\documentclass[preprint,12pt]{elsarticle}
\biboptions{numbers,sort&compress}

\usepackage{eepic}
\usepackage{amsmath,amssymb,amsthm}
\usepackage[suffix=, outdir=./]{epstopdf}
\usepackage{multicol}

\usepackage{float}
\usepackage{caption}
\usepackage{subfig}

\newtheorem{theorem}{Theorem}

\begin{document}
	\begin{frontmatter}
		
		\title{Lyapunov dimension formula \\ of attractors in the Tigan and Yang systems}
		
		\author{G. A. Leonov}
		
		\author{N. V. Kuznetsov\corref{cor}}
		\ead{nkuznetsov239@gmail.com}
		
		\author{N. A. Korzhemanova}
		
		\author{D. V. Kusakin}
		
		
		\address[spbu]{Faculty of Mathematics and Mechanics,
			St. Petersburg State University, 198504 Peterhof,
			St. Petersburg, Russia}
		
		\address[fin]{Department of Mathematical Information Technology,
			University of Jyv\"{a}skyl\"{a}, \\ 40014 Jyv\"{a}skyl\"{a}, Finland}

\begin{abstract}
    In latter days the technique of attractors dimension estimate
    of Lorenz type systems is actively developed.
	In this work the Lyapunov dimension of attractors of the Tigan and Yang systems is estimated.
\end{abstract}

		
\end{frontmatter}
	
	\section{Introduction}
	Nowadays various estimates of Lyapunov dimension of attractors of
	generalized Lorenz systems
	\begin{equation}\label{genlorenz}
		\left\{
		\begin{aligned}
			&\dot x= \sigma (y - x),\\
			&\dot y= r x - d y - xz,\\
			&\dot z= -b z + xy,
		\end{aligned}
		\right.
	\end{equation}
	are actively developed  \cite{Leonov-2012-PMM,LeonovK-2015-AMC}.
    One particular case of \eqref{genlorenz} system is the Yang system	\cite{YangChen-2008}: 
	\begin{equation}\label{sys:Yang}
		\left\{
		\begin{aligned}
			&\dot x= \sigma(y-x),\\
			&\dot y= r x - xz,\\
			&\dot z=-b z+xy,
		\end{aligned}
		\right.
	\end{equation}
	where $\sigma, b$ are positive, $r$ is arbitrary real number.
	In \cite{ChenY-2013} an estimation of Lyapunov dimension
    for the generalized Lorenz system \eqref{genlorenz} was obtained (see also remarks in \cite{LeonovKKK-2015-arXiv-LuChen}),
    which for the case $d=0$ (i.e. for the Yang system \eqref{sys:Yang}) is as follows:
 	\begin{theorem}\label{ErrorTheorem} \cite{ChenY-2013}
		Let $K$ be an invariant compact set of the Yang system \eqref{sys:Yang}.
		Assume $\sigma > 0$, $b > 0$, $r > 0$, $b(\sigma - b) + \sigma r > 0$, and if one of the following conditions holds
		\begin{enumerate}
			\item $2 \sigma \ge b$, $\sigma^2 (3b + r) - b^3 - \sigma b(6b+r) >0$
			\item $2\sigma < b$, $3\sigma^2 (b+r) - 6 \sigma b^2 - b^3 >0$,
		\end{enumerate}
		then
		$\dim_L K \le 3 - \frac{2 (\sigma + b)}{\sigma + \sqrt{\sigma^2 + 4 \sigma r}}$.
	\end{theorem}
	
In our work we extend the domain of parameters, where the above estimation is valid.

\section{Lyapunov functions in the dimension theory}\label{sec:1}
Consider an autonomous differential equation
\begin{equation}
\dot{{\bf \rm x}} = {\bf \rm f}({\bf \rm x}), \quad
{\bf \rm x} \in \mathbb{R}^n,
\label{eq:ode}
\end{equation}
where ${\bf \rm f} : \mathbb{R}^n \to \mathbb{R}^n$ is  smooth.
Suppose that solution ${\bf \rm x}(t, {\bf \rm x}_0),\ x(0,x_0)=x_0$ exist for all $t\geq0$
and consider corresponding linearized system along the solution:
\begin{equation}\label{eq:lin_eq}
	\dot{{\bf \rm u}} = J({\bf \rm x}(t, {\bf \rm x}_0)) \, {\bf \rm u},
	\quad {\bf \rm u} \in \mathbb{R}^n, ~t \in \mathbb{R}_+,
\end{equation}
where
\[
J({\bf \rm x}(t, {\bf \rm x}_0)) =
\left[
\frac{\partial f_i({\bf \rm x})}{\partial {\bf \rm x}_j}\big|_{{\bf \rm x}
	= {\bf \rm x}(t, {\bf \rm x}_0)}
\right]
\]
is the $(n \times n)$ Jacobian matrix.
By a linear variable change
${\bf\rm y} = S {\bf\rm x}$ with a nonsingular $(n \times n)$-matrix $S$
system \eqref{eq:ode} is transformed into
\[
\dot{{\bf\rm y}} = S \,\dot{{\bf\rm x}} = S \,{\bf\rm f}
(S^{-1}{\bf\rm y}) = \tilde{{\bf\rm f}}({\bf\rm y}).
\]
Consider the linearization along the corresponding solution
${\bf\rm y}(t, {\bf\rm y}_0) = S {\bf\rm x}(t, S^{-1} {\bf\rm x}_0)$:
\begin{equation}\label{eq:lin_eq-new}
  \dot {\bf \rm v} = \tilde{J}({\bf \rm y}(t, {\bf \rm y}_0))\,{\bf \rm v},
  \quad {\bf \rm v} \in \mathbb{R}^n.
\end{equation}
Here the Jacobian matrix is as follows
\begin{align}
\tilde{J}({\bf\rm y}(t, {\bf\rm y}_0))
=S \, J({\bf\rm x}(t, {\bf\rm x}_0)) \, S^{-1}. \label{jacobian-new}
\end{align}

Suppose that
$\lambda_1 ({\bf \rm x},S) \geqslant \cdots \geqslant \lambda_n ({\bf \rm x},S)$
are eigenvalues of the matrix
\begin{equation}
\frac{1}{2}
\left( S \, J({\bf\rm x}(t, {\bf\rm x}_0)) \, S^{-1} +
(
S \, J({\bf\rm x}(t, {\bf\rm x}_0)) \, S^{-1}
)^{*}\right).
\label{SJS}
\end{equation}

\begin{theorem}[\cite{Leonov-2002,Leonov-2012-PMM}]\label{theorem:th1}
	Given an integer $j \in [1,n-1]$ and $s \in [0,1]$,
	suppose that there are a continuously
	differentiable scalar function $\vartheta: \mathbb{R}^n \rightarrow \mathbb{R}$
	and a nonsingular matrix $S$ such that
	\begin{equation}\label{ineq:th-1}
	\lambda_1 ({\bf \rm x},S) + \cdots + \lambda_j ({\bf \rm x},S) + s\lambda_{j+1}
	({\bf \rm x},S) + \dot{\vartheta}({\bf \rm x}) < 0,
	~ \forall \, {\bf \rm x} \in K.
	\end{equation}
	Then $\dim_L K \leqslant j+s$.
\end{theorem}
Here $\dot{\vartheta}$ is the derivative of $\vartheta$ with respect
to the vector field ${\bf\rm f}$:
$$
\dot{\vartheta} ({\bf \rm x}) = ({\rm grad}(\vartheta))^{*}{\bf\rm f}({\bf \rm
	x}).
$$
The introduction of the matrix $S$ can be regarded as a change of the space metric.

\begin{theorem}[\cite{Leonov-1991-Vest,LeonovB-1992,BoichenkoLR-2005,Leonov-2012-PMM}]
	\label{theorem:th2}
	Assume that there are a continuously differentiable scalar function $\vartheta$
	and a nonsingular matrix $S$ such that
	\begin{equation}\label{ineq:th-2}
		\lambda_1 ({\bf \rm x},S) + \lambda_2 ({\bf \rm x},S) +
		\dot{\vartheta}({\bf \rm x}) < 0,
		~ \forall \, {\bf \rm x} \in \mathbb{R}^n.
	\end{equation}
	Then any solution of system \eqref{eq:ode} bounded on $[0,+\infty)$
	tends to an equilibrium as $t \rightarrow +\infty$.
\end{theorem}
Thus, if \eqref{ineq:th-2} holds,
then the global attractor of system \eqref{eq:ode}
coincides with its stationary set.

\section{Main result}
We can prove the following result for the Yang system.
\begin{theorem}\label{Main}
	
	\begin{enumerate}
		\item Assume $r = 0$ and the following inequalities $b(\sigma - b) > 0$, $\sigma - \frac{(\sigma + b)^2}{4(\sigma - b)} \ge 0$ are satisfied.
		Then any bounded on $[0; +\infty)$ solution of system \eqref{sys:Yang}
		tends to a certain equilibrium as $t \to +\infty$.
		
		\item Assume $r < 0$ and $ r \sigma + b(\sigma - b) > 0$.
		Then any bounded on $[0; +\infty)$ solution of system \eqref{sys:Yang}
		tends to a certain equilibrium as $t \to +\infty$.
		
		\item Assume $r > 0$ and there are two distinct real roots $\gamma^{(II)} > \gamma^{(I)}$ of equation
		\begin{equation}\label{eq:mainTheorem:rPos}
			4 b r \sigma^2 (\gamma + 2\sigma - b)^2 + 16 \sigma b \gamma (r \sigma^2 + b(\sigma + b)^2 - 4 \sigma (\sigma r + \sigma b - b^2)) =0,
		\end{equation}
		such that $\gamma^{(II)} > 0$.
		
		In this case
		\begin{enumerate}
			\item if
			\begin{equation*}\label{cond:mainTheorem:rPosStable}
				b(b - \sigma) < r \sigma < b (\sigma + b),
			\end{equation*}
			then any bounded on $[0; +\infty)$ solution of system \eqref{sys:Yang}
			tends to a certain equilibrium as $t \to +\infty$.
			
			\item if
			\begin{equation}\label{cond:mainTheorem:sPosFormula}
				r \sigma > b (\sigma + b),
			\end{equation}
			then
			\begin{equation}\label{formula}
				\dim_L K = 3 - \frac{2 (\sigma + b)}{\sigma + \sqrt{{\sigma}^2 + 4 \sigma r}},
			\end{equation}
			where $K \supset (0,0,0)$ is bounded invariant set of system \eqref{sys:Yang}.
		\end{enumerate}
	\end{enumerate}  	
\end{theorem}

For proving Theorem \ref{Main} we use Theorems~\ref{theorem:th1} and \ref{theorem:th2} with the matrix $S$ of the form
\begin{equation*}
	S = \begin{pmatrix}
	-\rho^{-1} & 0 & 0\\
	-\frac{b}{\sigma} & 1 & 0\\
	0 & 0 & 1
	\end{pmatrix},
\end{equation*}
 and the function
 \begin{equation*}
	 \vartheta (x, y, z) = \frac{(1-s) V(x, y, z)}{[{\sigma}^2+4\sigma r]^{\frac{1}{2}}},
 \end{equation*}
 where
$$\rho = \frac{\sigma}{\sqrt{\sigma r + b(\sigma - b)}},$$
$$V(x, y, z) = \gamma_4 x^2+(\sigma\gamma_2+\gamma_3)y^2+\gamma_3z^2-\frac{1}{4\sigma}\gamma_2 x^4+\gamma_2 x^2 z + \gamma_1 \gamma_2 xy -\frac{\sigma}{b}z.$$
 The parameters $\gamma_1, \gamma_2, \gamma_3, \gamma_4$ are variable.

Exact value of the Lyapunov dimension is obtained
by the comparison of estimate \eqref{ineq:th-2}
and the Lyapunov dimension at the zero equilibrium point.		

We consider classical chaotic parameters of the Yang system: $\sigma = 10, b = \frac{8}{3}, r = 16$ and
$\sigma = 35, b =3, r = 35$. In the first case inequality
\eqref{cond:mainTheorem:sPosFormula} is as follows
$$ 160 > \frac{304}{9}$$
and the greater root of corresponding equation \eqref{eq:mainTheorem:rPos} is equal
$$\gamma_1 ^{(II)} = \frac{6358}{135}+\frac{14}{135}\sqrt{178309}.$$
In the second case inequality \eqref{cond:mainTheorem:sPosFormula} is as follows
$$ 1225 > 114$$
and the greater root of \eqref{eq:mainTheorem:rPos} is equal
$$\gamma_1 ^{(II)} =\frac{193391}{1225}+\frac{2}{1225}\sqrt{7665943314}.$$
Thus, classical chaotic parameters satisfy the requirement of Theorem \ref{Main} and the formula \eqref{formula} is valid.

\bigskip
\section{T-system}
Consider the Tigan system (T-system) \cite{Tigan-2008,JiangHB-2010}:
\begin{equation}\label{T-sys}
	\left\{
	\begin{aligned}
		&\dot x= a(y-x),\\
		&\dot y= (c-a) x - axz,\\
		&\dot z=-b z+xy.
	\end{aligned}
	\right.
\end{equation}
By the transformation
$$x \rightarrow \frac{x}{\sqrt{a}}, y \rightarrow \frac{y}{\sqrt{a}}, z \rightarrow \frac{z}{a}$$
one has
\begin{equation}
	\left\{
	\begin{aligned}
		&\dot x= a(y-x)\\
		&\dot y= (c-a) x - xz\\
		&\dot z=-b z+xy.
	\end{aligned}
	\right.
\end{equation}
So the T-system can be transformed to the Yang system with the following parameters $\sigma = a, r = c-a$.
Both systems were independently considered in 2008 year.
But a particular case of the T-system was considered in 2004 \cite{Tigan-2004}.

The T-system has classical chaotic parameters $(a, b, c) = (2.1, 0.6, 30)$.
Conditions of Theorem~\ref{Main} are valid in the case
$\sigma = \frac{21}{10}, b = \frac{6}{10}, r = \frac{279}{10}$ and takes the form:

(1) parameter $r=27.9 > 0$,

(2) inequality \eqref{cond:mainTheorem:sPosFormula} is transformed to
$$\frac{5859}{100} > \frac{81}{50},$$
and, thus, is valid,

(3) the roots of the corresponding equation \eqref{eq:mainTheorem:rPos} are $\gamma_1^{(I)} = \frac{9883}{1085}-\frac{1}{1085}\sqrt{82416853}$ and $\gamma_1^{(II)} = \frac{9883}{1085}+\frac{1}{1085}\sqrt{82416853}$,
and, thus, the greater root is positive.

Therefore, if $K \supset (0,0,0)$ is a bounded invariant set of system \eqref{T-sys}
with the parameters $\sigma = 2.1, b = 0.6, r = 27.9$, then
$$\dim_L K = 3 - \frac{2 (\sigma+b)}{\sigma + \sqrt{\sigma^2 + 4\sigma r}} \approx 2.692345984.$$

\section{Conclusions}
The Yang system \eqref{sys:Yang} by the transformation
\begin{equation} \label{transformationToTwoParam}
	x \rightarrow \sigma x, y\rightarrow \sigma y, z \rightarrow \sigma z, t \rightarrow \frac{t}{\sigma}, \quad \sigma \neq 0
\end{equation}
takes the form
\begin{equation}
	\left\{
	\begin{aligned}
		&\dot x= y-x,\\
		&\dot y= \frac{r}{\sigma} x - xz,\\
		&\dot z=-\frac{b}{\sigma} z+xy.
	\end{aligned}
	\right.
\end{equation}
For $\sigma=0$ the Yang system becomes linear and its dynamics has minor interest.
Thus, without loss of generality, one can assume that $\sigma=1$.
Below we compare the domains of parameters in Theorem~\ref{Main} and Theorem~\ref{ErrorTheorem} for $\sigma=1$.
	
\begin{figure}[!ht]
	\centering
	\subfloat[]{
		\label{fig:yangdomain}
		\includegraphics[width=0.34\textwidth, angle=270]{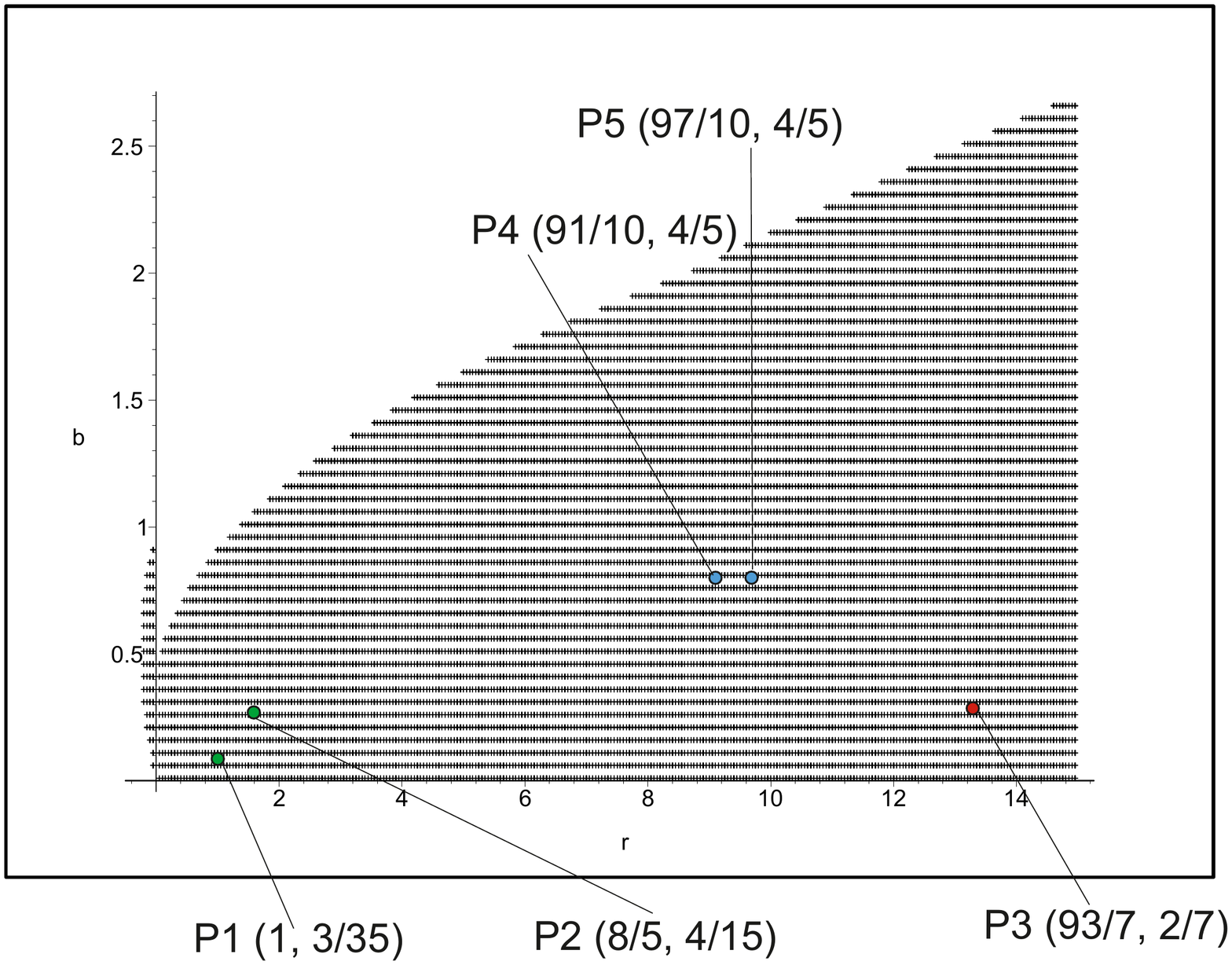}
	}~
	\subfloat[]{
		\label{fig:chinedomain}
		\includegraphics[width=0.34\textwidth, angle=270]{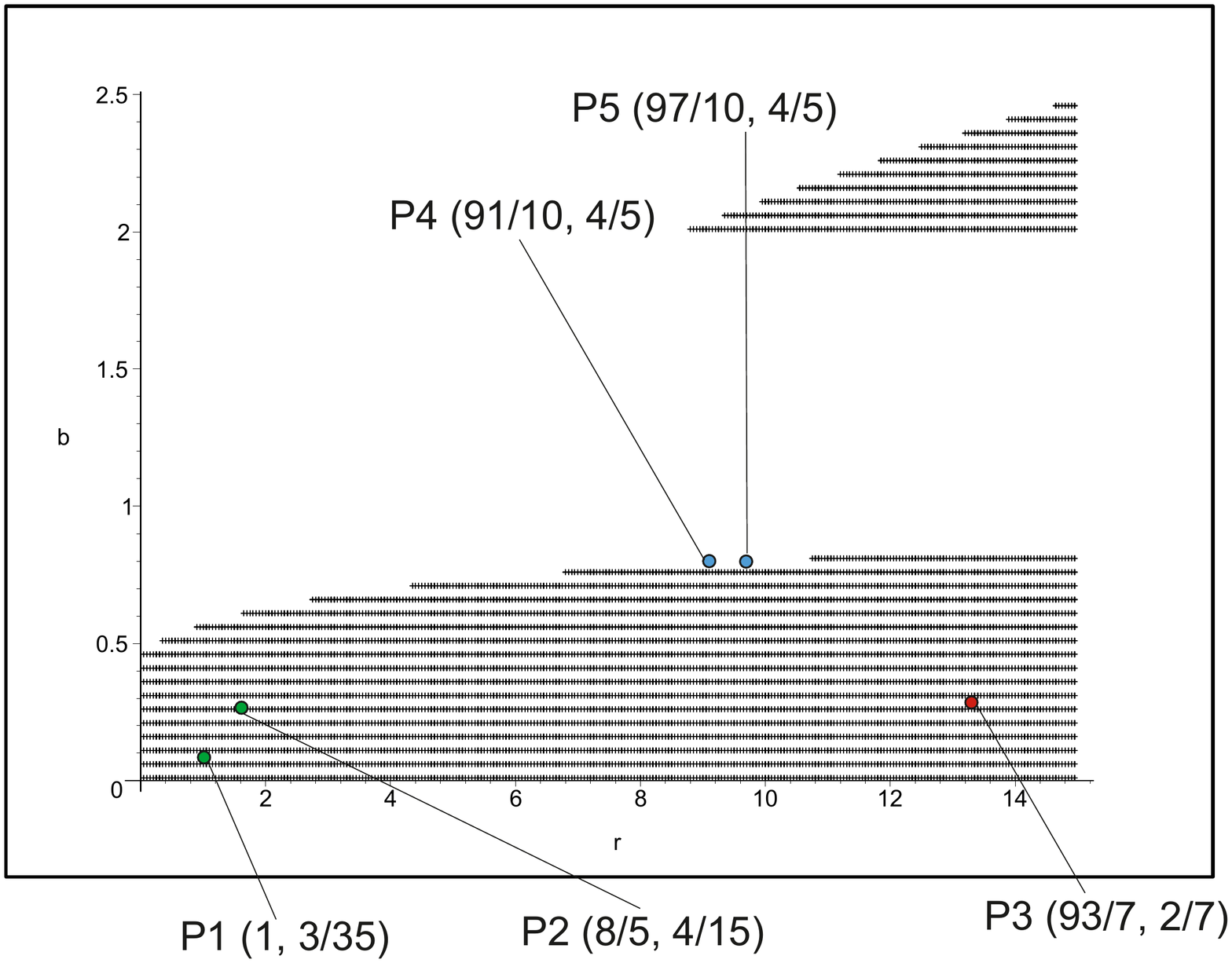}
	}
	\caption{Domains of parameters for $\sigma=1,\ r < 15$.
		(a)  Theorem~\ref{Main}.
		(b) Theorem~\ref{ErrorTheorem}.}
\end{figure}

The points $P1$ and $P2$ are the parameters which correspond
the classical chaotic self-excited attractors\footnote{
An attractor is called a self-excited attractor
if its basin of attraction
intersects with any open neighborhood of an equilibrium,
otherwise it is called a hidden attractor
\cite{LeonovKV-2011-PLA,LeonovKV-2012-PhysD,LeonovK-2013-IJBC,LeonovKM-2015-EPJST}.
 Up to now in such Lorenz-like systems as
Lorenz, Chen, Lu and Tigan systems
 self-excited chaotic attractors have been found only.
 In such Lorenz-like systems as Glukhovsky--Dolghansky and Rabinovich systems
 both self-excited and hidden attractors can be found
\cite{KuznetsovLM-2015,LeonovKM-2015-CNSNS,LeonovKM-2015-EPJST}.
Recent examples of hidden attractors can be found in
\emph{The European Physical Journal Special Topics: Multistability: Uncovering Hidden Attractors}, 2015
(see \cite{ShahzadPAJH-2015-HA,BrezetskyiDK-2015-HA,JafariSN-2015-HA,ZhusubaliyevMCM-2015-HA,SahaSRC-2015-HA,Semenov20151553,FengW-2015-HA,Li20151493,FengPW-2015-HA,Sprott20151409,Pham20151507,VaidyanathanPV-2015-HA}).
Note that while coexisting self-excited attractors
can be found by the standard computational procedure,
there is no regular way to predict the existence
or coexistence of hidden attractors.
}  in the Yang system,
$P3$ --- classical chaotic self-excited attractor in the T-system,
$P4, P5$ --- parameters, for which we have also found chaotic self-excited attractors by numerical analysis.

\begin{figure}[!ht]
	\centering
	\subfloat[]{
		\label{fig:newyang1}
		\includegraphics[width=0.52\textwidth]{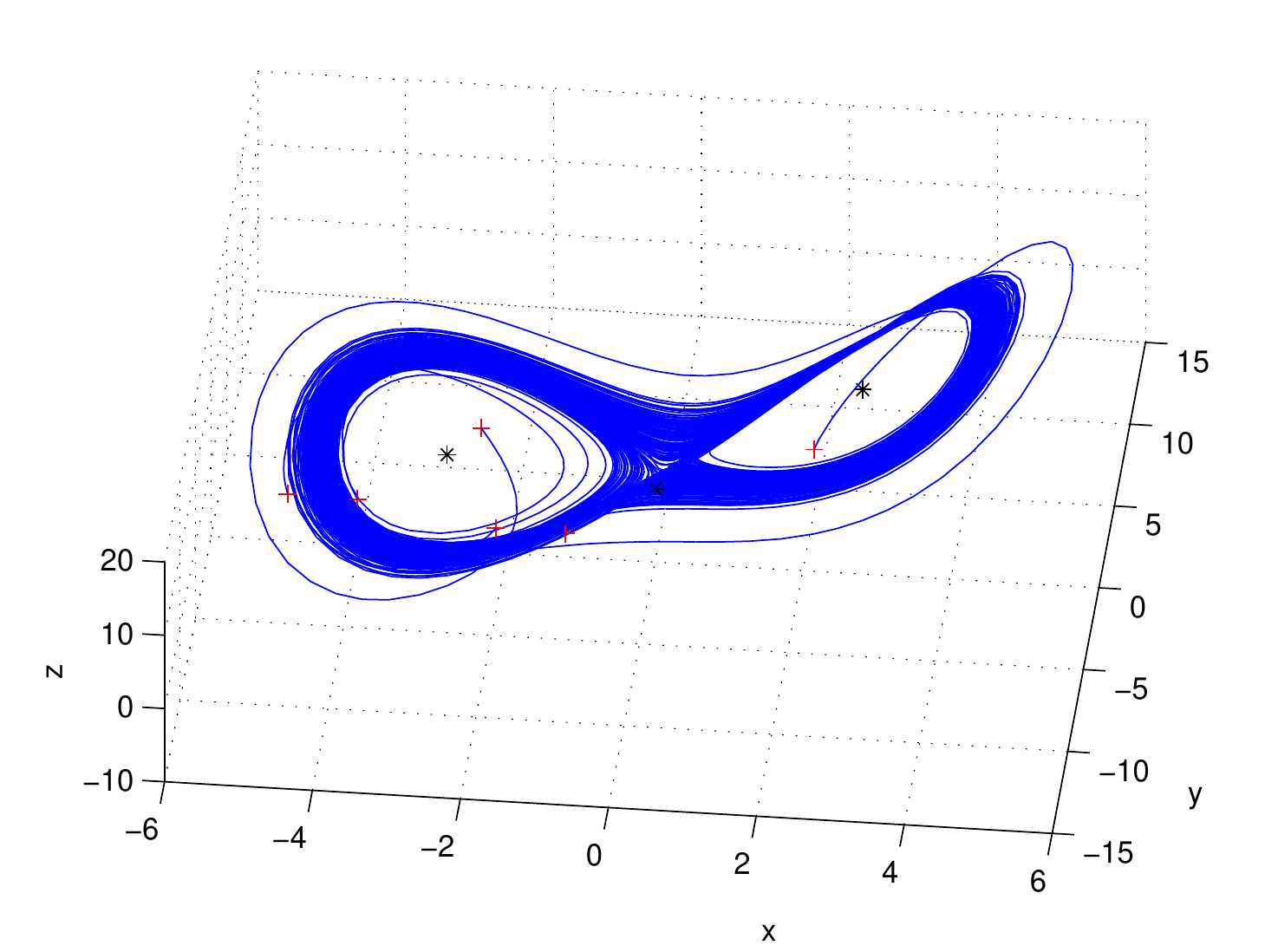}
	}~
	\subfloat[]{
		\label{fig:newyang2}
		\includegraphics[width=0.52\textwidth]{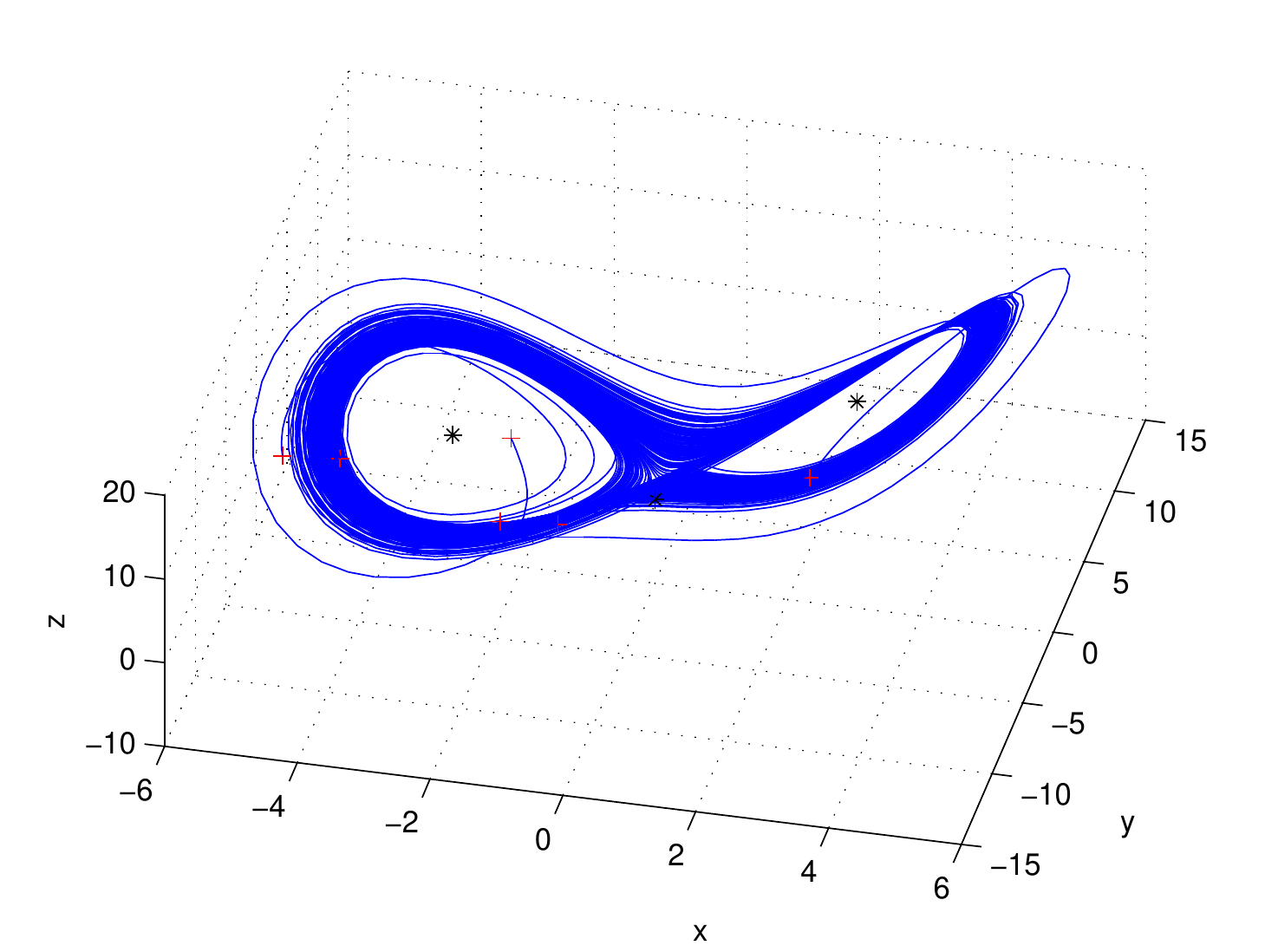}
	}
	\caption{
        Self-excited attractors in the Yang system:
		(a) P4 --- $\sigma$ = 1, b = 0.8, r = 9.1.
		(b) P5 --- $\sigma$ = 1, b = 0.8, r = 9.7.}
\end{figure}

\bigskip


\end{document}